# A Multivariate Variance Components Model for Analysis of Covariance in Designed Experiments

**James G. Booth, Walter T. Federer, Martin T. Wells and Russell D. Wolfinger**

*Abstract.* Traditional methods for covariate adjustment of treatment means in designed experiments are inherently conditional on the observed covariate values. In order to develop a coherent general methodology for analysis of covariance, we propose a multivariate variance components model for the joint distribution of the response and covariates. It is shown that, if the design is orthogonal with respect to (random) blocking factors, then appropriate adjustments to treatment means can be made using the univariate variance components model obtained by conditioning on the observed covariate values. However, it is revealed that some widely used models are incorrectly specified, leading to biased estimates and incorrect standard errors. The approach clarifies some issues that have been the source of ongoing confusion in the statistics literature.

## 1. INTRODUCTION

This article concerns the adjustment of treatment means in designed experiments to account for one or more covariates. Analysis of covariance has a long history, dating back to Fisher (1934). Much of its development in the context of designed experiments followed soon after. Examples can be found in many classical textbooks, including Federer (1955),

*James Booth e-mail:* Jim.Booth@cornell.edu *and Martin Wells e-mail:* mtw1@cornell.edu *are Professors and Walter Federer e-mail:* wtf1@cornell.edu *was Emeritus Professor, Department of Biological Statistics and Computational Biology, Comstock Hall, Cornell University, Ithaca, New York 14853, USA. Russell Wolfinger is Director of Scientific Discovery and Genomics, SAS Institute Inc., Cary, North Carolina 27513, USA e-mail:* Russ.Wolfinger@jmp.com.

Cochran and Cox (1957), Snedecor and Cochran (1967), as well as more recent texts such as Milliken and Johnson (2002). We are specifically interested in settings in which the covariates are random variables, not fixed by the design. Also, we generally suppose that the covariate values are not affected by the treatments, for example, because they were measured prior to application of the treatments. The multivariate mixed model we propose can be modified to handle problems in which this assumption is not valid. However, there are additional inferential issues in this case.

As a canonical example we discuss in detail the randomized complete blocks (RCB) design with a single covariate. In the classical analysis of this design the blocks are treated as fixed, and the covariate is included as a predictor. The least squares treatment means obtained from the fixed blocks model fit adjust the arithmetic treatment means to account for differences in the average covariate measurements among the treatments. Including the covariate block means as an additional predictor has no effect on the least squares fit because differences at the block level are already accounted for. Treating blocks as fixed effectively confines the scope of inference to only those blocks and covariate values in the study. In particular, the standard errors obtained from the fixed effects model for the least squares treatment means do not account for repeated sampling of blocks.

However, in most applications the blocks in the study can be viewed as a random sample from a population of interest, and it is of interest to extend the scope of inference to the population of blocks. An obvious modification of the classical model in this case is to simply treat the block effects as random; that is, to fit a (univariate) mixed model with two variance components. Under the standard independence and normality assumptions, adjusted treatment means obtained from the mixed model have the same form as the classical least squares means, except that the covariate regression coefficient is a weighted average of the estimates obtained from intra- and inter-block regressions.

A key point made in this paper is that both univariate fixed and mixed models for analysis of covariance are inherently conditional on the measured covariate values. An obvious question is, therefore, what joint distribution for response and covariate leads to the conditional models. We show that by simply treating the block effects as random in the randomized complete blocks design setting, one is implicitly assuming that the marginal block variance for the covariate is zero; that is, an implied model for the joint distribution that is not realistic. As a result, the adjusted treatment means obtained from the naive, univariate mixed model are biased. However, by starting with a bivariate variance components model for the joint distribution of response and covariate, one is led to a sensible univariate conditional model, which properly accounts for the design with respect to the covariate. The idea of a bivariate model is suggested in Cox and McCullagh [(1982), Section 7], but not fully developed. Multivariate variance components models are discussed in Khuri, Mathew and Sinha [(1998), Chapter 10], but the application to analysis of covariance is not considered. The fully conditional approach was also advocated by Neuhaus and McCulloch (2006) who consider the situation where random effects in a generalized linear mixed model may be correlated with one of the predictors. Classical likelihood approaches lead to inconsistent estimators in this setting. Results in Neuhaus and McCulloch (2006) show that conditional maximum likelihood can eliminate the bias.

An outline of the remainder of the paper is as follows. In the next section we look back historically and attempt to explain why some fundamental issues in analysis of covariance are still unresolved. The randomized complete blocks design is discussed in detail in Section 3. In Section 4 we show how the bivariate model for the randomized complete blocks design can be generalized to orthogonal designs, and to allow adjustment for multiple covariates. In the orthogonal case the conditional model for the response given the covariates implied by the multivariate mixed model for the joint distribution turns out to be a univariate mixed model. This implies that appropriate adjustment of treatment means can be accomplished using standard software. The methodology is applied to some standard examples of orthogonal designs. The nonorthogonal case is discussed in Section 5. Here we show that appropriate adjustment cannot be accomplished by fitting a univariate mixed model. However, an EM algorithm for fitting a general multivariate linear variance components model is developed using arguments that parallel those for the univariate case, as discussed in Searle, Casella and McCulloch (1992), Chapter 8. The methodology is applied to balanced incomplete block designs and unbalanced designs. We conclude with some discussion in Section 6.

## 2. HISTORICAL PERSPECTIVE

Since analysis of covariance in designed experiments has such a long history, readers may wonder why confusion over such basic modeling issues persists even today. We attempt to explain this by discussing the topic in its historical context. Arguably, the heyday of analysis of covariance was pre-1960, predating the widespread use of matrix algebra in statistics and clearly before the availability of high-speed computing. Matrices are two hundred and some years old but their use in statistics only became commonplace in the late 1950s. The very first paper in the first issue of the *Annals of Mathematical Statistics* by Wichsell (1930) is entitled "Remarks on Regression," yet it has no matrices. It is likely that the lateness of adoption of matrices arose from their being treated as a topic in pure mathematics and, hence, their practical use in statistical modeling remained hidden. In the classic design books Cochran and Cox (1957) and Federer (1955), there is no mention of matrices, whereas in Kempthorne (1952) the design problem is formulated as a general linear model but is not applied to analyze any advanced designs. Kempthorne [(1952), page 66] even notes that, at the time, there did not appear to be a complete matrix formulation of the general linear model anywhere in the literature. In unbalanced data it was a longtime puzzle why statistical methods gave two different least squares estimates of fixed effects in a one-way classification depending upon whether one assumed that one effect was zero, or that all the effects summed to zero. The literature had certainly not kept up with R. C. Bose's (1949) concept of estimability. Nor were 1950s design and linear model researchers aware that Penrose's (1955) generalized inverse matrix could be used to solve the normal equations in the nonfull rank setting, as in design problems. With the aid of a generalized inverse, Rao (1962) demonstrated how the unique unbiased estimators of estimable functions could be constructed. In a recent *Statistical Science* conversation (Wells (2009)) Shayle Searle points out that random effects modeling in the 1950s was quite limited and mostly for balanced data. One of the early formulations of matrix methods in variance and covariance components analysis can be found in Searle (1956).

An excellent paper by Zelen (1957) reviews the thinking at the time for balanced incomplete block (BIB) designs. The algebraic manipulations associated with the multivariate model described in this paper would be very difficult, if not impossible, without the use of matrix algebra and facility with the multivariate normal distribution, mathematical tools that were not fully developed in the statistics literature at the time of Zelen's paper. We focus on Zelen's discussion of intra- and inter-block regressions, which we summarize with the following two quotes from Section 3 of his paper:

> in the analysis of covariance, the inter-block model will be important if the variability of the concomitant variable is large for "between blocks" as compared to the variability "within blocks." This situation may allow more precise inter-block estimates of the regression coefficients as compared to the corresponding intra-block estimates and later, when discussing the slopes of intra- and inter-block regressions,

> Some statisticians, however, have advocated a more general model which allows the intra-block regression coefficients to be different from the inter-block regression. In this paper, all models are such that the intra-block regression is the same as that for the inter-block regression. It is difficult for this writer to visualize situations allowing separate regressions.

It turns out that the "[s]ome statisticians" Zelen referred to were right, but why? Clearly their arguments were not entirely convincing at the time. The bivariate variance components model described in Section 3.3 reveals that Zelen's two statements are incompatible. In fact, between block variation in the covariate implies that the intra- and inter-block regressions are not the same. Putting it another way, forcing the two slopes to be equal amounts to an assumption that there is no variation among the block covariate means. This assumption is clearly violated in practical circumstances and therefore leads to biased (adjusted) treatment means as well as inconsistent estimates of variance components.

## 3. RANDOMIZED COMPLETE BLOCKS DESIGN

### 3.1 Classical Approach

Consider a randomized complete blocks design with response, $Y$, and associated concomitant variable, $Z$. Suppose that $Z$ is measured prior to application of the treatment but is possibly correlated with the response. Let $i = 1, \ldots, t$ be the index for treatment, and $j = 1, \ldots, b$ be the index for block. The classical fixed effects linear model for this design is as follows:

$$\text{(1)} \qquad Y_{ij} = \mu + \tau_i + \beta_j + \gamma z_{ij} + E_{ij},$$

where $E_{ij} \sim N(0, \sigma_e^2)$, independently, and $\sum_i \tau_i = \sum_j \beta_j = 0$. Notice that replacing the covariate $z_{ij}$ with the block centered value $z_{ij} - \bar{z}_{.j}$ has no effect on the fit of this model because the term, $-\gamma \bar{z}_{.j}$, can be incorporated into the fixed effect for block $j$. The adjusted mean for treatment $i$ is the estimated mean response at a fixed value of $Z$, conventionally taken to be its average observed value, $\bar{z}_{..}$.

Throughout this paper Greek letters represent fixed effects (unknown parameters), upper case Roman letters are random variables or known matrices, and lower case Roman letters are either observed values of random variables or known constants (or vectors).

Table 1
*Adjusted treatment means and standard errors for Pearce's apple yield data*

| Treatment | Univariate fixed | | Univariate mixed | | Bivariate mixed | |
|---|---|---|---|---|---|---|
| | Adj.Mean | Std.Err. | Adj.Mean | Std.Err. | Adj.Mean | Std.Err. |
| A | 280.48 | 6.37 | 280.41 | 13.69 | 280.48 | 12.98 |
| B | 266.57 | 6.36 | 266.55 | 13.68 | 266.57 | 12.98 |
| C | 274.07 | 6.36 | 274.05 | 13.68 | 274.07 | 12.98 |
| D | 281.14 | 6.44 | 281.32 | 13.72 | 281.14 | 13.02 |
| E | 300.92 | 6.72 | 301.33 | 13.87 | 300.92 | 13.19 |
| S | 251.34 | 6.86 | 250.85 | 13.95 | 251.34 | 13.28 |

Adjusted means and their standard errors for the apple yield data from Pearce (1953, 1982), based on fixed effects, univariate mixed effects and bivariate mixed effects models. The standard errors involve the ML estimates of variance components.

With these conventions it is implicit in the notation that model (1) characterizes the conditional distribution of the response given the observed values of the concomitant variable.

We define the inter-block regression model to be the implied model for the block means, specifically,

$$\bar{Y}_{\cdot j} = \mu + \beta_j + \gamma \bar{z}_{\cdot j} + \bar{E}_{\cdot j}.$$

Thus, in this context, the inter-block model contains no information about treatment differences, or about the regression parameter, $\gamma$, because the terms, $\gamma \bar{z}_{\cdot j}$, are confounded with the block effects. We define the intra-block regression model using the $t-1$ orthogonal Helmert contrasts, $\mathbf{h}_2, \ldots, \mathbf{h}_t$, between components of the observation vector, $\mathbf{Y}_j$, for block $j$. Specifically, let $Y_{ij}^* = \mathbf{h}_i^T \mathbf{Y}_j$, for $i = 2, \ldots, t$ and $j = 1, \ldots, b$, and similarly define $z_{ij}^*$ and $E_{ij}^*$. Then the intra-block model in this case is

$$Y_{ij}^* = \tau_i^* + \gamma z_{ij}^* + E_{ij}^*,$$

where $\tau_i^* = \mathbf{h}_i^T \boldsymbol{\tau}$, $i = 2, \ldots, t$, are $t-1$ orthogonal contrasts among the treatment means. It follows that all the information about treatment differences, and the covariate regression parameter, is contained in the intra-block model.

The adjusted treatment means are estimates of the mean responses when $Z = \bar{z}_{\cdot\cdot}$. These are given by

$$(2) \quad \hat{\mu}_{i,\text{adj}} = \hat{\mu} + \hat{\tau}_i + \hat{\gamma}\bar{z}_{\cdot\cdot} = \bar{y}_{i\cdot} - \hat{\gamma}(\bar{z}_{i\cdot} - \bar{z}_{\cdot\cdot}),$$

for $i = 1, \ldots, t$, where $\hat{\gamma}$ is the BLUE for $\gamma$. These do not involve the (estimated) block effects because they are averages over the blocks and the block effects sum to zero. In the fixed effects case, $\hat{\gamma}$ is the

ordinary least squares estimate

$$(3) \quad \hat{\gamma}_{ols} = \frac{\mathbf{z}^T(\mathbf{C}_t \otimes \mathbf{C}_b)\mathbf{y}}{\mathbf{z}^T(\mathbf{C}_t \otimes \mathbf{C}_b)\mathbf{z}},$$

where $\mathbf{C}_t = \mathbf{I}_t - \bar{\mathbf{J}}_t$ is the centering matrix of dimension $t$, and $\mathbf{y} = (y_{11}, y_{12}, \ldots, y_{tb})^T$ is the entire response vector, with $\mathbf{z}$ defined analogously. Since $\mathbf{C}_b \bar{\mathbf{J}}_b = 0$, it follows from (3) that $\hat{\gamma}_{ols}$ is independent of the unadjusted treatment mean vector, $(\mathbf{I}_t \otimes \bar{\mathbf{J}}_b)\mathbf{y}$, with components, $\bar{y}_{i\cdot}$, $i = 1, \ldots, t$. Hence, the variance formula for the adjusted means based on the traditional model with fixed block effects is

$$(4) \quad \text{var}(\hat{\mu}_{i,\text{adj}}) = \frac{\sigma_e^2}{b} + \frac{\sigma_e^2}{\mathbf{z}^T(\mathbf{C}_t \otimes \mathbf{C}_b)\mathbf{z}}(\bar{z}_{i\cdot} - \bar{z}_{\cdot\cdot})^2.$$

For numerical illustration we consider the apple yield data from Pearce (1953, 1982). In this experiment there were $b = 4$ blocks, and $t = 6$ treatments (A, B, C, D, E and S), with S being the standard practice in English apple orchards of keeping the land clean in the summer. The response, $Y$, is the yield per plot, and the covariate, $Z$, is the number of boxes of fruit, measured to the nearest tenth of a box, for the four seasons previous to the application of the treatments. The adjusted treatment means and their estimated standard errors, based on three different models, are reported in Table 1.

### 3.2 Univariate Mixed Model

In most applications the blocks can be regarded as a random sample from a population, and it is of interest to make inferences about the average treatment effects across the population of potential blocks. In such cases it makes sense to treat the block effects as random. Thus, model (1) becomes

$$(5) \quad Y_{ij} = \mu + \tau_i + B_j + \gamma z_{ij} + E_{ij},$$

where now $B_j \sim$ i.i.d. $N(0, \sigma_b^2)$ independently of the random errors, $E_{ij}$. Replacing the covariate $z_{ij}$ with the data centered value, $z_{ij} - \bar{z}..$, has no effect on the fit of this model because the term, $-\gamma\bar{z}..$, can be incorporated into the fixed intercept. However, unlike the fixed block model (1), replacing the covariate with block centered values, $z_{ij} - \bar{z}.j$, does affect the fit.

The inter-block regression model derived from (5) is

$$(6) \qquad \bar{Y}.j = \mu + \gamma\bar{z}.j + B_j + \bar{E}.j,$$

while the intra-block model is the same as in the fixed effects case. Thus, when the block effects are treated as random, they are incorporated into the error term of the inter-block model. As a result, the inter-block model does contain additional information about the covariate regression parameter. In particular, it would appear from (6) that the information in the inter-block model will increase with the variability of the covariate block means. This explains the first quote from Zelen (1957) given in Section 2 (albeit for a BIB design).

The adjusted treatment means based on model (5) have the form (2), being the expected treatment means in repeated sampling (involving different blocks) at a common concomitant variable value, $Z = \bar{z}..$. However, the BLUE for $\gamma$ is a weighted average of the estimates obtained from the intra- and inter-block regression models, where the weights are inversely proportional to their variances. Let $\rho$ denote correlation between any two sample treatment means, that is,

$$\rho = \text{cor}(\bar{Y}_i, \bar{Y}_k) = \frac{\sigma_b^2}{\sigma_b^2 + \sigma_e^2/t}.$$

Then, it is shown in the Appendix that the BLUE of $\gamma$ based on (5) has the form

$$(7) \qquad \begin{aligned} \hat{\gamma}_{\text{mixed}} &= \frac{\mathbf{z}^T[(\mathbf{I}_t - \rho\bar{\mathbf{J}}_t) \otimes \mathbf{C}_b]\mathbf{y}}{\mathbf{z}^T[(\mathbf{I}_t - \rho\bar{\mathbf{J}}_t) \otimes \mathbf{C}_b]\mathbf{z}} \\ &= \frac{\mathbf{z}^T(\mathbf{C}_t \otimes \mathbf{C}_b)\mathbf{y} + (1-\rho)\mathbf{z}^T(\bar{\mathbf{J}}_t \otimes \mathbf{C}_b)\mathbf{y}}{\mathbf{z}^T[(\mathbf{I}_t - \rho\bar{\mathbf{J}}_t) \otimes \mathbf{C}_b]\mathbf{z}}. \end{aligned}$$

If the block variance dominates the error variance, and hence $\rho \approx 1$, then the mixed effects estimate of $\gamma$ is close to the ordinary least squares estimate in (3). On the other hand, if the block variance is dominated by the error variance, then $\rho \approx 0$, and the estimate in (7) corresponds to the fixed effect case in which the block effects are omitted from the model.

The adjusted means and their standard errors based on the mixed effects model (5), for the apple yield data, are tabulated in Table 1. The ML variance component estimates in this example are $\hat{\sigma}_e^2 = 194.55$ and $\hat{\sigma}_b^2 = 553.98$, resulting in a correlation estimate $\hat{\rho} = 0.9447$, and $\hat{\gamma} = 28.89$. This compares with the ordinary least squares estimate $\hat{\gamma} = 28.40$. Thus, in this case the adjusted mean values are quite similar. However, the standard errors reported by the software are quite different. This is because inferences from the fixed effects model (1) are restricted to the four blocks in the study, whereas those from the model (5) apply to the population of blocks. Specifically, since (7) implies $\hat{\gamma}_{\text{mixed}}$ is independent of the unadjusted treatment means,

$$\begin{aligned} &\text{var}(\hat{\mu}_{i,\text{adj}}) \\ &= \frac{\sigma_e^2 + \sigma_b^2}{b} \\ &\quad + \frac{\mathbf{z}^T[(\mathbf{I}_t - \rho\bar{\mathbf{J}}_t) \otimes \mathbf{C}_b]\mathbf{\Sigma}[(\mathbf{I}_t - \rho\bar{\mathbf{J}}_t) \otimes \mathbf{C}_b]\mathbf{z}}{(\mathbf{z}^T[(\mathbf{I}_t - \rho\bar{\mathbf{J}}_t) \otimes \mathbf{C}_b]\mathbf{z})^2} \\ &\quad \cdot (\bar{z}_i - \bar{z}..)^2, \end{aligned}$$

where $\mathbf{\Sigma} \equiv \text{var}(\mathbf{Y}) = \sigma_e^2 \mathbf{I}_t \otimes \mathbf{I}_b + \sigma_b^2 \mathbf{J}_t \otimes \mathbf{I}_b$. Notice that, even as $\rho$ approaches 1, this variance formula still differs from the fixed effects variance given in (4) by an additive amount, $\sigma_b^2/b$, which accounts for variation due to sampling of blocks.

### 3.3 Bivariate Mixed Model

As noted in the Introduction, the models (1) and (5) are inherently conditional on the observed values of the covariate $Z$. The fixed effects model (1) is appropriate if the blocks in the experiment are the only ones of interest, whereas model (5) is an attempt to broaden the applicability of inferences to the hypothetical population from which the blocks were drawn. An obvious question is what model(s) for the joint distribution of $(Y, Z)$ leads to the conditional model (5)?

Consider a bivariate model in which the distribution of $Z$ is independent of the treatments but allows for random variation between blocks and residual error. Specifically,

$$(8) \qquad \begin{aligned} \begin{pmatrix} Y_{ij} \\ Z_{ij} \end{pmatrix} &= \begin{pmatrix} \mu_y \\ \mu_z \end{pmatrix} + \begin{pmatrix} \tau_{i,y} \\ 0 \end{pmatrix} \\ &\quad + \begin{pmatrix} B_{j,y} \\ B_{j,z} \end{pmatrix} + \begin{pmatrix} E_{ij,y} \\ E_{ij,z} \end{pmatrix} \end{aligned}$$

where the block effects are i.i.d. bivariate normal,

$$\begin{pmatrix} B_{j,y} \\ B_{j,z} \end{pmatrix} \sim \text{ i.i.d. } N_2 \left[ \begin{pmatrix} 0 \\ 0 \end{pmatrix}, \mathbf{\Sigma}_B = \begin{pmatrix} \sigma_{b,y}^2 & \sigma_{byz} \\ \sigma_{b,zy} & \sigma_{b,z}^2 \end{pmatrix} \right],$$

independently of the bivariate residual errors,

$$\begin{pmatrix} E_{ij,y} \\ E_{ij,z} \end{pmatrix} \sim \text{ i.i.d. } N_2 \left[ \begin{pmatrix} 0 \\ 0 \end{pmatrix}, \mathbf{\Sigma}_E = \begin{pmatrix} \sigma_{e,y}^2 & \sigma_{eyz} \\ \sigma_{e,zy} & \sigma_{e,z}^2 \end{pmatrix} \right].$$

As before, let $\mathbf{Y}_j = (Y_{1j}, \ldots, Y_{tj})^T$ denote the response vector for the $j$th block, and similarly define $\mathbf{Z}_j$. Then the conditionally specified model implied by bivariate model (8) can be formally derived using the fact that

$$
\begin{aligned}
(9) \quad & \begin{pmatrix} \mathbf{Y}_j \\ \mathbf{Z}_j \end{pmatrix} \\
& \sim \text{ i.i.d. } N_{2t} \left[ \begin{pmatrix} \boldsymbol{\mu}_y \\ \mu_z \mathbf{1}_t \end{pmatrix}, \begin{pmatrix} \sigma_{e,y}^2 \mathbf{I}_t + \sigma_{b,y}^2 \mathbf{J}_t \\ \sigma_{e,zy} \mathbf{I}_t + \sigma_{b,zy} \mathbf{J}_t \end{pmatrix} \right. \\
& \qquad\qquad \left. \begin{matrix} \sigma_{e,yz} \mathbf{I}_t + \sigma_{b,yz} \mathbf{J}_t \\ \sigma_{e,z}^2 \mathbf{I}_t + \sigma_{b,z}^2 \mathbf{J}_t \end{matrix} \right) \right],
\end{aligned}
$$

where $\boldsymbol{\mu}_y$ is the vector of treatment means with components, $\mu_{i,y} = \mu_y + \tau_{i,y}$. It follows that

$$
\begin{aligned}
(10) \quad & E(\mathbf{Y}_j | \mathbf{Z}_j = \mathbf{z}_j) \\
& = \boldsymbol{\mu}_y + (\sigma_{e,yz} \mathbf{I}_t + \sigma_{byz} \mathbf{J}_t)(\sigma_{e,z}^2 \mathbf{I}_t + \sigma_{b,z}^2 \mathbf{J}_t)^{-1} \\
& \qquad \cdot (\mathbf{z}_j - \mathbf{1}_t \mu_z) \\
& = \boldsymbol{\mu}_y + (\sigma_{e,yz} \mathbf{I}_t + t\sigma_{byz} \bar{\mathbf{J}}_t) \frac{1}{\sigma_{e,z}^2} \\
& \qquad \cdot \left( \mathbf{I}_t - \frac{t\sigma_{b,z}^2}{\sigma_{e,z}^2 + t\sigma_{b,z}^2} \bar{\mathbf{J}}_t \right)(\mathbf{z}_j - \mathbf{1}_t \mu_z) \\
& = \boldsymbol{\mu}_y + \gamma_e(\mathbf{z}_j - \mathbf{1}_t \mu_z) + \gamma_b \mathbf{1}_t(\bar{z}_{\cdot j} - \mu_z),
\end{aligned}
$$

where $\gamma_e = \sigma_{e,yz}/\sigma_{e,z}^2$ and

$$(11) \qquad \gamma_b = \frac{\sigma_{e,z}^2 \sigma_{b,yz} - \sigma_{e,yz} \sigma_{b,z}^2}{\sigma_{e,z}^2 (\sigma_{b,z}^2 + \sigma_{e,z}^2/t)}.$$

The conditional variance is

$$
\begin{aligned}
(12) \quad & \text{var}(\mathbf{Y}_j | \mathbf{Z}_j = \mathbf{z}_j) \\
& = \sigma_{e,y}^2 \mathbf{I}_t + \sigma_{b,y}^2 \mathbf{J}_t \\
& \quad - (\sigma_{e,yz} \mathbf{I}_t + \sigma_{byz} \mathbf{J}_t)(\sigma_{e,z}^2 \mathbf{I}_t + \sigma_{b,z}^2 \mathbf{J}_t)^{-1} \\
& \quad \cdot (\sigma_{e,zy} \mathbf{I}_t + \sigma_{b,zy} \mathbf{J}_t) \\
& = \sigma_{e,y}^2 \mathbf{I}_t + \sigma_{b,y}^2 \mathbf{J}_t \\
& \quad - (\sigma_{e,yz} \mathbf{I}_t + \sigma_{byz} \mathbf{J}_t) \frac{1}{\sigma_{e,z}^2}
\end{aligned}
$$

$$
\begin{aligned}
& \quad \cdot \left( \mathbf{I}_t - \frac{\sigma_{b,z}^2}{\sigma_{e,z}^2 + t\sigma_{b,z}^2} \mathbf{J}_t \right)(\sigma_{e,zy} \mathbf{I}_t + \sigma_{b,zy} \mathbf{J}_t) \\
& = \sigma_e^2 \mathbf{I}_t + \sigma_b^2 \mathbf{J}_t,
\end{aligned}
$$

where $\sigma_e^2 = \sigma_{e,y}^2 - \sigma_{e,yz}^2/\sigma_{e,z}^2$, and

$$(13) \qquad \sigma_b^2 = \sigma_{b,y}^2 - [\gamma_e \sigma_{b,yz} + \gamma_b(\sigma_{b,yz} + \sigma_{e,yz}/t)].$$

It follows from (10) and (12) that the univariate conditional model implied by (9) is

$$(14) \qquad Y_{ij} = \mu + \tau_i + B_j + \gamma_e z_{ij} + \gamma_b \bar{z}_{\cdot j} + E_{ij},$$

where $\mu = \mu_y - (\gamma_e + \gamma_b)\mu_z$, $\tau_i \equiv \tau_{i,y}$, and $B_j \sim$ i.i.d. $N(0, \sigma_b^2)$ independently of $E_{ij} \sim$ i.i.d. $N(0, \sigma_e^2)$. The inter-block regression model implied by (14) is

$$\bar{Y}_{\cdot j} = \mu + \gamma_{be} \bar{z}_{\cdot j} + B_j + \bar{E}_{\cdot j},$$

where

$$\gamma_{be} \equiv \gamma_e + \gamma_b = \frac{\sigma_{b,yz} + \sigma_{e,yz}/t}{\sigma_{b,z}^2 + \sigma_{e,z}^2/t}$$

is the slope of the inter-block regression. Thus, in this case the inter-block model contains no information about the intra-block covariate regression coefficient. Similarly, in the case of a generalized linear mixed model, where random effects may be correlated with one of the predictors, it is shown in Neuhaus and McCulloch (2006) that conditional maximum likelihood also leads naturally to the partitioning of the covariate into between- and within-cluster components.

Writing (13) in terms of the intra-block and inter-block slopes, $\gamma_e$ and $\gamma_{be}$, we obtain

$$\sigma_b^2 = \sigma_{b,y}^2 - \gamma_e \sigma_{e,yz}/t - \gamma_{be}(\sigma_{b,yz} + \sigma_{e,yz}/t).$$

Thus, the block variance for the response in the conditional model is the marginal block variance adjusted for intra- and inter-block regression on the covariate.

The univariate mixed model (5) is the conditional model implied by (8) when $\gamma_b = 0$, which only happens if $\sigma_{b,z}^2 = 0$, an unrealistic assumption in practice. At this point it is interesting to recall Zelen's (1957) comments, quoted earlier, concerning the equality of slopes in the inter- and intra-block models, and the information in the inter-block model about the intra-block slope increasing with the block-to-block variability in the covariate. It is now clear that these statements are incompatible. Block-to-block variability in the covariate implies that the inter- and intra-block slopes are different. For this reason the use of the univariate mixed model (5) leads to biased estimates of adjusted means and inconsistent estimates of variance components as the number of blocks increases.

We define the adjusted treatment means to be the expected responses if the covariate values were all equal to the average observed covariate value. Thus, the model (14) implies

$$
\begin{aligned}
(15) \qquad \mu_{i,\text{adj}} &= \mu + \tau_i + (\gamma_b + \gamma_e)\bar{z}.. \\
&= \mu_{i,y} + \gamma_{be}(\bar{z}.. - \mu_z).
\end{aligned}
$$

It is shown in the Appendix that $\hat{\mu}_z = \bar{z}..$, that $\hat{\gamma}_e$ equals the ordinary least squares estimate based on univariate fixed effects model, and that $\hat{\mu}_{i,y} = \bar{y}_{i\cdot} - \hat{\gamma}_e(\bar{z}_{i\cdot} - \bar{z}..)$. It follows that the adjusted treatment means based on (14) are identical to (2).

Estimates of the adjusted treatment means for the apple yield data based on (14) are given in Table 1. The estimate of the inter-block slope in this case is $\hat{\gamma}_{be} = 37.25$. This is quite different in magnitude (although not statistically) from the estimated intra-block slope, $\hat{\gamma}_e = 28.40$. Since the intra-block estimate is identical to those based on the univariate fixed effects model, the standard errors for the adjusted means are given by

$$
\text{var}(\hat{\mu}_{i,\text{adj}}) = \frac{\sigma_e^2 + \sigma_b^2}{b} + \frac{\sigma_e^2}{\mathbf{z}^T(\mathbf{C}_t \otimes \mathbf{C}_b)\mathbf{z}}(\bar{z}_{i\cdot} - \bar{z}..)^2.
$$

The estimated standard errors are larger than those based on the fixed effects model by an additive factor of $\sigma_b^2/b$. This is as is should be, because the scope of inference has been broadened to the population of blocks.

Up to now we have assumed that the covariate values are not affected by the treatments. If they are, then the bivariate model (8) is no longer appropriate. An obvious modification of (8) in this case is

$$
\begin{pmatrix} Y_{ij} \\ Z_{ij} \end{pmatrix} = \begin{pmatrix} \mu_y \\ \mu_z \end{pmatrix} + \begin{pmatrix} \tau_{i,y} \\ \tau_{i,z} \end{pmatrix} + \begin{pmatrix} B_{j,y} \\ B_{j,z} \end{pmatrix} + \begin{pmatrix} E_{ij,y} \\ E_{ij,z} \end{pmatrix}.
$$

The conditional model for $Y$ given $Z$ implied by this model has exactly the same form as (14). However, the treatment effect parameter $\tau_i$ is equal to $\tau_{i,y} - \gamma_e\tau_{i,z}$. This makes sense in that what is being estimated is the direct effect of treatments on the response mean, as opposed to the indirect effect through the covariate. As noted by Bartlett (1936), there is reason for caution in this setting due to hidden extrapolation. Comparing conditional expectations of treatment means at equal covariate levels may not make sense if the treatments affect what covariate values are observed.

## 4. GENERAL ORTHOGONAL BLOCKING DESIGNS

### 4.1 Theory

Let $\mathbf{Z}_{ij} = (Z_{ij1}, \ldots, Z_{ijm})^T$ be a covariate vector associated with the response $Y_{ij}$, for $i = 1, \ldots, k$ in replicate $j = 1, \ldots, b$. Thus, the data matrix for replicate $j$ is given by

$$
\begin{bmatrix} Y_{1j} & \mathbf{Z}_{1j}^T \\ Y_{2j} & \mathbf{Z}_{2j}^T \\ \vdots & \vdots \\ Y_{kj} & \mathbf{Z}_{kj}^T \end{bmatrix} = [\mathbf{Y}_j, \mathbf{Z}_j],
$$

say. Let $\mathbf{Z}_{jr}^*$ denote the $r$th column of $\mathbf{Z}_j$. Suppose that $\mathbf{Y}_j$ can be decomposed into $\mathbf{Y}_j = \boldsymbol{\mu}_y + \mathbf{T}_j + \mathbf{U}_j$, where $\boldsymbol{\mu}_y$ is the fixed mean of $\mathbf{Y}_j$, which depends on the treatments, $\mathbf{T}_j$ is the sum random factors associated with treatments (and therefore independent of $\mathbf{Z}_j$), and $\mathbf{U}_j$ is the sum of $q$ random design factors plus residual errors.

We suppose that vec$[\mathbf{U}_j, \mathbf{Z}_j]$, $j = 1, \ldots, b$, are i.i.d. multivariate normal vectors of dimension $k(m + 1)$, with means, vec$[\mathbf{0}_k, \boldsymbol{\mu}_z^T \otimes \mathbf{1}_k]$, where the components of $\boldsymbol{\mu}_z$ are the marginal means of the $m$ covariates, and with covariance matrix, $\mathbf{V}$. We say that the design is an "orthogonal blocking design" if the matrix $\mathbf{V}$ has the following structure. Let $\mathbf{A}_0 \equiv \bar{\mathbf{J}}_k$, and $\mathbf{A}_l$, $l = 1, \ldots, q$, be $k \times k$ matrices with the properties that (a) $\mathbf{A}_l$ is idempotent, (b) $\mathbf{A}_l\mathbf{A}_{l'} = 0$ if $l \neq l'$, and $\sum_{l=0}^{q} \mathbf{A}_l = \mathbf{I}_k$. Then we suppose there exist nonsingular matrices, $\mathbf{G}_0, \mathbf{G}_1, \ldots, \mathbf{G}_q$, each of dimension $m + 1$, such that

$$
(16) \qquad \mathbf{V} = \sum_{l=0}^{q} \mathbf{G}_l \otimes \mathbf{A}_l.
$$

Notice that a design can be orthogonal, in this sense, regardless of the assignment of the treatments.

In general, the matrix $\mathbf{V}$ is a function of $(q + 1)\frac{1}{2}(m + 1)(m + 2)$ free variance and covariance parameters which determine the $q+1$ variance-covariance matrices, $\boldsymbol{\Sigma}_0, \ldots, \boldsymbol{\Sigma}_q$, associated with the residual errors, and the $q$ random design factors. In particular, we note that the variance–covariance structure for $\mathbf{U}_j$ is

$$
\mathbf{V}_{uu} = \sum_{l=0}^{q} g_{l,uu}\mathbf{A}_l,
$$

where $g_{l,uu}$, $l = 0, \ldots, q$, are scalar parameters, and that this structure is that implied by orthogonality of the random blocking factors.

Example. Consider the RCB design discussed in Section 3. In this case there is only one covariate, so $m = 1$. The vector $\mathbf{U}_j$ associated with the $j$th block consists of the sum of the block effect and and the residual error vector,

$$\mathbf{U}_j = \mathbf{1}_t B_j + \mathbf{E}_j.$$

Finally, the covariance matrix for $(\mathbf{Y}_j^T, \mathbf{Z}_j^T)^T$ is given by

$$\mathbf{V} = \mathbf{\Sigma}_E \otimes \mathbf{I}_t + \mathbf{\Sigma}_B \otimes \mathbf{J}_t$$
$$= (\mathbf{\Sigma}_E + t\mathbf{\Sigma}_B) \otimes \bar{\mathbf{J}}_t + \mathbf{\Sigma}_E \otimes (\mathbf{I} - \bar{\mathbf{J}}_t).$$

The fact that $\mathrm{vec}[\mathbf{U}_j, \mathbf{Z}_j]$, $j = 1, \ldots, b$, are i.i.d. multivariate normal vectors implies that marginally $\mathrm{vec}(\mathbf{Z}_j)$, $j = 1, \ldots, b$, are i.i.d. $N(\boldsymbol{\mu}_z \otimes \mathbf{1}_k, \mathbf{V}_{zz})$, where $u$ and $z$ subscript combinations are used to denote components of the partitioned matrix. Moreover, conditionally upon $\mathbf{Z}$, $\mathbf{U}_j$, $j = 1, \ldots, b$, have independent normal distributions with means

$$E(\mathbf{U}_j | \mathbf{Z}_j = \mathbf{z}_j)$$
$$= \mathbf{V}_{uz} \mathbf{V}_{zz}^{-1} (\mathbf{z}_j - \boldsymbol{\mu}_z \otimes \mathbf{1}_k)$$
$$= \left( \sum_{l=0}^q \mathbf{g}_{l,uz} \otimes \mathbf{A}_l \right) \left( \sum_{l=0}^q \mathbf{G}_{l,zz}^{-1} \otimes \mathbf{A}_l \right) (\mathbf{z}_j - \boldsymbol{\mu}_z \otimes \mathbf{1}_k)$$
$$= \left( \sum_{l=0}^q \mathbf{g}_{l,uz} \mathbf{G}_{l,zz}^{-1} \otimes \mathbf{A}_l \right) (\mathbf{z}_j - \boldsymbol{\mu}_z \otimes \mathbf{1}_k)$$
$$= \left( \sum_{l=0}^q \boldsymbol{\gamma}_l^T \otimes \mathbf{A}_l \right) (\mathbf{z}_j - \boldsymbol{\mu}_z \otimes \mathbf{1}_k),$$

where $\boldsymbol{\gamma}_l^T = \mathbf{g}_{l,uz} \mathbf{G}_{l,zz}^{-1}$ is a $1 \times m$ parameter vector. Since $(\boldsymbol{\gamma}_l^T \otimes \mathbf{A}_l)(\boldsymbol{\mu}_z \otimes \mathbf{1}_k) = \boldsymbol{\gamma}^T \boldsymbol{\mu}_z \otimes \mathbf{A}_l \mathbf{1}_k = \mathbf{0}$, unless $l = 0$, in which case it equals $\boldsymbol{\gamma}_0^T \boldsymbol{\mu}_z \mathbf{1}_k$, we have

$$E(\mathbf{U}_j | \mathbf{Z}_j = \mathbf{z}_j) = -\boldsymbol{\gamma}_0^T \boldsymbol{\mu}_z \mathbf{1}_k + \sum_{l=0}^q \sum_{r=1}^m \gamma_{lr} \mathbf{A}_l \mathbf{z}_{jr}^*.$$

Finally, since $\mathbf{T}_j$ has mean zero, and is independent of $\mathbf{Z}_j$,

$$E(\mathbf{Y}_j | \mathbf{Z}_j = \mathbf{z}_j) = \boldsymbol{\mu}_y^c + \sum_{l=0}^q \sum_{r=1}^m \gamma_{lr} \mathbf{A}_l \mathbf{z}_{jr}^*,$$

where $\boldsymbol{\mu}_y^c = \boldsymbol{\mu}_y - \boldsymbol{\gamma}_0^T \boldsymbol{\mu}_z \mathbf{1}_k$. Thus, the conditional mean of the response is given by a linear model with treatment effects incorporated into $\boldsymbol{\mu}_y^c$, and covariate regression effects with slopes, $\{\gamma_{lr}\}$, $l = 0, \ldots, q$, associated with each of the $m$ covariates, $r = 1, \ldots, m$. Since $\mathbf{A}_l \mathbf{1}_k = \mathbf{0}$ for $l > 0$, the expected response if all

the covariates are equal to their respective marginal means is

$$\boldsymbol{\mu}_{\mathrm{adj}} = \boldsymbol{\mu}_y - \mathbf{1}_k \boldsymbol{\gamma}_0^T (\bar{\mathbf{z}}_{..}^* - \boldsymbol{\mu}_z),$$

which generalizes the formula (15) for the RCB design.

The conditional variance of $\mathbf{U}_j$ is given by

$$\mathrm{var}(\mathbf{U}_j | \mathbf{Z}_j = \mathbf{z}_j) = \mathbf{V}_{uu} - \mathbf{V}_{uz} \mathbf{V}_{zz}^{-1} \mathbf{V}_{zu}$$
$$= \sum_{l=0}^q (g_{l,uu} - \mathbf{g}_{l,uz} \mathbf{G}_{l,zz}^{-1} \mathbf{g}_{l,zu}) \otimes \mathbf{A}_l$$
$$= \sum_{l=0}^q \lambda_l \mathbf{A}_l,$$

corresponding to an orthogonal design with orthogonal partition $\{\mathbf{A}_l\}$.

Example. Consider again the RCB design of Section 3. Note that the conditional mean (10) can be reexpressed in the form,

$$E(\mathbf{Y}_j | \mathbf{Z}_j = \mathbf{z}_j) = \boldsymbol{\mu}_y^c + \gamma_{be} \bar{\mathbf{J}}_t \mathbf{z}_j + \gamma_e (\mathbf{I}_t - \bar{\mathbf{J}}_t) \mathbf{z}_j,$$

where $\boldsymbol{\mu}_y^c = \boldsymbol{\mu}_y - \gamma_{be} \mu_z$, and $\gamma_{be} = \gamma_e + \gamma_b$. Moreover, the conditional variance (12) can be reexpressed as

$$\mathrm{var}(\mathbf{Y}_j | \mathbf{Z}_j = \mathbf{z}_j) = (\sigma_e^2 + t\sigma_b^2) \bar{\mathbf{J}}_t + \sigma_e^2 (\mathbf{I}_t - \bar{\mathbf{J}}_t).$$

## 4.2 Examples

4.2.1 *Split-plot designs.* Consider a standard split-plot experiment with $t$ whole-plot treatments, each replicated $r$ times, and $s$ split-plot treatments in each wholeplot. Let $Y_{ijk}$ denote response to split-plot treatment $k$, in whole-plot $j$ assigned to wholeplot treatment $i$. Similarly index the covariate values $Z_{ijk}$. Then, the marginal models for the response and covariate are

$$Y_{ijk} = \mu_y + \alpha_{i,y} + W_{(i)j,y} + \tau_{k,y} + \alpha\tau_{ik,y} + E_{ijk,y}$$

and

$$Z_{ijk} = \mu_z + W_{(i)j,z} + E_{ijk,z}$$

respectively. Bivariate normality for the pairs, $(W_{(i)j,y}, W_{(i)j,z})$ and $(E_{ijk,y}, E_{ijk,z})$, imply that the conditional model for appropriate covariate adjustment has the form,

$$(17) \quad \begin{aligned} Y_{ijk} = \mu + \alpha_i + W_{(i)j} + \tau_k + \alpha\tau_{ik} \\ + \gamma_w \bar{z}_{ij.} + \gamma_e z_{ijk} + E_{ijk}, \end{aligned}$$

where $\alpha_i$ is the fixed main effect of the $i$th wholeplot treatment, $\tau_k$ is the fixed main effect of the $k$th split-plot treatment, and $W_{(i)j}$ is the random effect of the $j$th wholeplot replicate nested within the $i$th wholeplot treatment. Milliken and Johnson [[2002], Section 15.4] discuss a split-plot design in the context of a cookie baking experiment in which oven temperature is the whole plot factor, and cookie type is the split-plot factor. The covariate in their example is the thickness of the slices of cookie dough, but their proposed "equal slopes" model does not include the whole plot regression term in (17).

If the experiment is arranged in $b$ blocks, with $r$ replicate wholeplots for each wholeplot treatment level in each block, then the marginal model for the response is

$$Y_{ijkl} = \mu_y + B_{i,y} + \alpha_j + (B\alpha)_{ij} + W_{(ij)k,y}$$
$$+ \tau_l + (B\tau)_{il} + (\alpha\tau)_{jl} + (B\alpha\tau)_{ijl} + E_{ijkl,y}.$$

Since the treatments have no effect on the covariate, the marginal model for $Z$ is

$$Z_{ijkl} = \mu_z + B_{i,z} + W_{(ij)k,z} + E_{ijkl,z}.$$

Notice that, in this case, there are random interactions between the blocking factor and treatments that affect the response, but not the covariate. Bivariate normality of the pairs, $(B_{i,y}, B_{i,z})$, $(W_{(ij)k,y}, W_{(ij)k,z})$ and $(E_{ijkl,y}, E_{ijkl,z})$, results in a conditional model for the response with a covariate adjustment at the individual response level, as well as adjustments for covariate variation in wholeplot and block means, specifically,

$$Y_{ijkl} = \mu + B_i + \alpha_j + (B\alpha)_{ij} + W_{(ij)k}$$
$$+ \tau_l + (B\tau)_{il} + (\alpha\tau)_{jl} + (B\alpha\tau)_{ijl}$$
$$+ \gamma_b \bar{z}_{i\cdots} + \gamma_w \bar{z}_{ijk\cdot} + \gamma_e z_{ijkl} + E_{ijkl}.$$

4.2.2 *Latin square design.* Let $Y_{ijk}$ denote the response in cell $(i, j)$ of a latin square design involving two random blocking factors and a fixed treatment factor, each with $b$ levels. An appropriate model ignoring any covariate information is

$$Y_{rijk} = \mu_y + R_{i,y} + C_{j,y} + \tau_k + E_{ijk,y}.$$

A marginal model for a random covariate is

$$Z_{rijk} = \mu_z + R_{i,z} + C_{j,z} + E_{ijk,z}.$$

Bivariate normality of the pairs, $(R_{i,y}, R_{i,z})$, $(C_{j,y}, C_{j,z})$ and $(E_{ijk,y}, E_{ijk,z})$, results in a conditional model for the response with a covariate adjustment at the individual response level, as well as adjustments for covariate variation in row and column means. That is,

$$Y_{ijk} = \mu + R_i + C_j + \tau_k + \gamma_r \bar{z}_{i\cdots}$$
$$+ \gamma_c \bar{z}_{\cdot j\cdot} + \gamma_e z_{ijk} + E_{ijk}.$$

4.2.3 *Incomplete block designs.* Consider an incomplete block design with $k < t$ treatments appearing in each block. Let $\mathbf{Y}_j$ and $\mathbf{Z}_j$ denote the response and covariate vectors for block $j$. The arguments of Section 3.3 lead to the conditional model (14), with the subscript $i$ taking $k$ values in $\{1, 2, \ldots, t\}$ depending on the value of $j$, and with the block regression parameter, $\gamma_b$, having the same form as (11) with $t$ replaced by $k$. We note here that even though this design is not orthogonal with respect to the response, it is orthogonal from the perspective of the covariate. It follows that the appropriate adjustment for covariates in an incomplete block design can be carried out using a univariate mixed model.

As an example we consider data from a study conducted by the National Bureau of Standards, discussed in Zelen [[1957], Section 6], to determine the effects of four geometrical shapes on the current noise of resistors. As described by Zelen, the "geometrical shapes were rectangular parallelepipeds (all having the same thickness) formed by taking all four combinations of 2 widths $(w_1, w_2)$ and 2 lengths $(l_1, l_2)$." Three resistors were mounted on each of 12 ceramic plates according to a BIB design. The response was the logarithm of the noise measurement, and the covariate was the logarithm of the resistance of each resistor. Estimated treatment effects and their standard errors obtained using the univariate mixed model of the form (5), and using the conditional model (14) derived from the bivariate model (8), are given in Table 2. There are substantial differences in both the estimated effects and the standard errors obtained using the two models. The estimates are also highly correlated, and these correlations must be taken into account in comparisons among the length and width combinations. In particular, Zelen considered the interaction contrast,

$$\pi = (l_2 w_1 - l_2 w_2) - (l_1 w_1 - l_1 w_2).$$

Estimates of $\pi$ under the two models (5) and (14) are 0.022 and 0.018 respectively, with standard errors 0.056 and 0.061. Thus, both models lead to the same conclusion that there is little statistical evidence for interaction.

## 5. NONORTHOGONAL DESIGNS

### 5.1 Factorization

A key feature of the multivariate mixed model in the orthogonal design case is that the parameters in the conditional model for $Y$ ($\boldsymbol{\mu}_y^c$ and $\boldsymbol{\gamma}_l$, $l =$

| Treatment | Univariate | | Bivariate | |
| | Effect | Std.Err. | Effect | Std.Err. |
|---|---|---|---|---|
| $l_1w_1$ | $-0.519$ | 0.112 | $-0.449$ | 0.233 |
| $l_1w_2$ | $-0.238$ | 0.029 | $-0.229$ | 0.040 |
| $l_2w_1$ | 0.249 | 0.031 | 0.238 | 0.045 |
| $l_2w_2$ | 0.508 | 0.109 | 0.440 | 0.226 |

Estimated treatment effects and standard errors obtained using the mixed models (5) and (14). In each case the variance components were estimated REML which explains why the first set of estimates (labeled "univariate") differ slightly from those obtained by Zelen (1957).

$0, 1, \ldots, q$) are variation independent of those in the marginal model for $Z$ ($\boldsymbol{\mu}_z$ and $\mathbf{G}_{l,zz}$, $l = 0, 1, \ldots, q$). The two sets of parameters combined represent a 1–1 transformation of the bivariate model parameterization ($\boldsymbol{\mu}_y$, $\boldsymbol{\mu}_z$, $\mathbf{G}_l$, $l = 0, 1, \ldots, q$). In general, this decomposition of the parameter space may not be possible, in which case appropriate adjustment of the treatment means cannot be accomplished using a univariate mixed model. To see this, suppose that the covariate data is only partially observed, say, $\mathbf{z} = (\mathbf{z}_o, \mathbf{z}_m)$, where $\mathbf{z}_o$ denotes the observed part, and $\mathbf{z}_m$ the unobserved. Then, the joint distribution of the data is

$$f(\mathbf{y}, \mathbf{z}_o; \boldsymbol{\mu}_y, \boldsymbol{\mu}_z, \mathbf{G})$$
$$= \int f_{Y|Z}(\mathbf{y}|\mathbf{z}; \boldsymbol{\mu}_y^c, \boldsymbol{\gamma}) f_Z(\mathbf{z}; \boldsymbol{\mu}_z, \mathbf{G}_{zz}) \, d\mathbf{z}_m,$$

and, hence, the marginal distribution of $\mathbf{z}_o$ is

$$\iint f_{Y|Z}(\mathbf{y}|\mathbf{z}; \boldsymbol{\mu}_y^c, \boldsymbol{\gamma}) f_Z(\mathbf{z}; \boldsymbol{\mu}_z, \mathbf{G}_{zz}) \, d\mathbf{z}_m \, d\mathbf{y}.$$

There is now no guarantee that the parameters that determine the marginal distribution of $\mathbf{z}_o$ will be separable from those that determine the conditional distribution of $\mathbf{y}$ given $\mathbf{z}_o$.

To further illustrate this point, consider again the RCB design discussed in Section 3. In this case the bivariate model has $t + 7$ parameters which determine the marginal means and block and error covariance matrices, ($\boldsymbol{\mu}_y, \mu_z, \boldsymbol{\Sigma}_B, \boldsymbol{\Sigma}_E$). The parameterization in terms of the marginal model for $Z$, and the conditional model for $Y$, is a union of two variation independent components of dimensions 3 and $t + 4$ respectively. Specifically, $(\mu_z, \sigma_{b,z}^2, \sigma_{e,z}^2) \cup (\boldsymbol{\mu}, \gamma_e, \gamma_b, \sigma_e^2, \sigma_b^2)$, where $\boldsymbol{\mu}$ has $i$th component equal to $\mu + \tau_i$. Now suppose that only $k < t$ covariate values are recorded in block $j$. Let $\mathbf{z}_{j,o}$ denote the observed vector of covariate values (of length $k$) and let $\bar{z}_{.j,o}$ denote its mean value. Then, modifying the arguments that led to equation (10) results in the conditional mean

$$E(\mathbf{Y}_j | \mathbf{Z}_{j,o} = \mathbf{z}_{j,o}) = \boldsymbol{\mu}_y + \gamma_e(\mathbf{z}_{j,o} - \mathbf{1}_k \mu_z)$$
$$+ \gamma_{b,o} \mathbf{1}_k(\bar{z}_{.j,o} - \mu_z),$$

where $\gamma_{b,o}$ has the same functional form as (11) with $t$ replaced by $k$. Thus, if the blocks have different numbers of covariate measurements, the parameters of the conditional model for $Y$ are not separable from those of the marginal model for $Z$.

## 5.2 General Multivariate Mixed Model

To simplify the notation, we relabel the vector of responses as $\mathbf{Z}_0$ (i.e., $\mathbf{Z}_0 \equiv \mathbf{Y}$). Then $\mathbf{Z} = \text{vec}[\mathbf{Z}_0, \mathbf{Z}_1, \ldots, \mathbf{Z}_m]$ is a vector containing all the responses and associated values of $m$ covariates stacked on top of one another. Thus, if the number of responses is $n$, then $\mathbf{Z}$ has length $n \times (m + 1)$. The multivariate mixed model described in this paper can be written in the form,

$$\mathbf{Z} = \mathbf{X}\boldsymbol{\beta} + \sum_{i=1}^{r} \mathbf{C}_i \mathbf{T}_i + \sum_{i=0}^{q} \mathbf{D}_i \mathbf{B}_i,$$

where $\mathbf{X}$ determines the means structure, $\mathbf{T}_i \sim N(\mathbf{0}, \sigma_i^2 \mathbf{I}_{c_i})$, independently for $i = 1, \ldots, r$, are random factors associated with treatments, and $\mathbf{B}_i \sim N(\mathbf{0}, \boldsymbol{\Sigma}_i \otimes \mathbf{I}_{d_i})$, independently for $i = 0, 1, \ldots, q$, are random (blocking) factors associated with the design, with the exception of $\mathbf{B}_0$, which is the residual error term (so that $d_0 \equiv n$). It is convenient to partition the matrices $\mathbf{X}$, $\mathbf{C}_i$ and $\mathbf{D}_i$ into blocks consisting of the $n$ rows associated with the response, or one of the $m$ covariates. Thus, $\mathbf{X} = [\mathbf{X}_0^T, \mathbf{X}_1^T, \ldots, \mathbf{X}_m^T]^T$, $\mathbf{C}_i = [\mathbf{C}_{i0}^T, \ldots, \mathbf{C}_{im}^T]^T$ and $\mathbf{B}_i = [\mathbf{B}_{i0}^T, \ldots, \mathbf{B}_{im}^T]^T$. Note that, if the covariates are unaffected by the treatments, then $\mathbf{C}_{ij} \equiv \mathbf{0}$ for $j > 0$. The model implies that $\mathbf{Z}$ has variance-covariance matrix equal to

$$\mathbf{V} \equiv \text{var}(\mathbf{Z}) = \sum_{i=1}^{r} \mathbf{C}_i \mathbf{C}_i^T \sigma_i^2 + \sum_{i=0}^{q} \mathbf{D}_i(\boldsymbol{\Sigma}_i \otimes \mathbf{I}_{d_i})\mathbf{D}_i^T.$$

We define the adjusted response mean vector as its conditional expectation given the covariates evaluated at their estimated mean values. If we partition the variance matrix, $\mathbf{V}$, into $n \times n$ matrix components, the conditional expectation of the response vector is given by

$$E(\mathbf{Y}|\mathbf{Z}_i = \mathbf{1}_n \hat{\mu}_{z_i}, i = 1, \ldots, m)$$
$$= \mathbf{X}_0 \boldsymbol{\beta} + [_r \mathbf{V}_{0i}]([\mathbf{V}_{ij}]_{i,j=1}^m)^{-1}[_c \mathbf{1}_n \hat{\mu}_{z,i} - \mathbf{1}_n \mu_{z,i}].$$

Here, we have used the notational definitions in Searle, Casella and McCulloch [[1992], Section 8.3]. Thus, for example, $[_r \mathbf{V}_{0i}] = [\mathbf{V}_{01}, \ldots, \mathbf{V}_{0m}]$. The estimate of the adjusted mean response vector is therefore

$$\hat{\boldsymbol{\mu}}_{\text{adj}} = \mathbf{X}_0 \hat{\boldsymbol{\beta}} = \mathbf{X}_0 (\mathbf{X}^T \hat{\mathbf{V}}^{-1} \mathbf{X})^{-1} \mathbf{X}^T \hat{\mathbf{V}}^{-1} \mathbf{Z}.$$

A "naive" variance-covariance formula for the adjusted mean vector, ignoring variability due to the estimation of $\mathbf{V}$, is given by the conditional variance of $\hat{\boldsymbol{\mu}}_{\text{adj}}$ assuming $\mathbf{V}$ is known. Specifically,

$$\text{var}(\hat{\boldsymbol{\mu}}_{\text{adj}}) = \mathbf{X}_0 (\mathbf{X}^T \mathbf{V}^{-1} \mathbf{X})^{-1} \mathbf{X}^T [\mathbf{V}^{i0} \mathbf{V}_{00}^* \mathbf{V}^{0j}]_{i,j=0}^m$$
$$\cdot \mathbf{X} (\mathbf{X}^T \mathbf{V}^{-1} \mathbf{X})^{-1} \mathbf{X}_0^T,$$

where $[\mathbf{V}^{ij}] = \mathbf{V}^{-1}$, and $\mathbf{V}_{00}^* = \mathbf{V}_{00} - [_r \mathbf{V}_{0i}] \cdot ([\mathbf{V}_{ij}]_{i,j=1}^m)^{-1}[_c \mathbf{V}_{0j}]$.

## 5.3 EM Algorithm

The distributional assumptions described above imply that the "complete" data vector,

$$(\mathbf{Z}^T, \mathbf{T}_1^T, \ldots, \mathbf{T}_r^T, \mathbf{B}_1^T, \ldots, \mathbf{B}_q^T)^T,$$

has a multivariate normal distribution with mean $(\boldsymbol{\beta}^T \mathbf{X}^T, \mathbf{0}^T)^T$. The assumptions imply the covariance between $\mathbf{Z}$ and $\mathbf{T}_i$ and $\mathbf{B}_i$ are, respectively,

$$\text{cov}(\mathbf{Z}, \mathbf{T}_i^T) = \mathbf{C}_i \sigma_i^2,$$

and

$$\text{cov}(\mathbf{Z}, \mathbf{B}_i^T) = \mathbf{D}_i (\boldsymbol{\Sigma}_i \otimes \mathbf{I}_{d_i}).$$

Thus, the joint density of the complete data vector is

$$f(\mathbf{z}, \mathbf{t}, \mathbf{b}) = |2\pi \boldsymbol{\Sigma}|^{-1/2} \exp(-Q/2),$$

where $Q = [(\mathbf{z} - \mathbf{X}\boldsymbol{\beta})^T, \mathbf{t}^T, \mathbf{b}^T] \boldsymbol{\Sigma}^{-1} [(\mathbf{z} - \mathbf{X}\boldsymbol{\beta})^T, \mathbf{t}^T, \mathbf{b}^T]^T$, and

$$\boldsymbol{\Sigma} = \begin{bmatrix} \mathbf{V} & \{_r \mathbf{C}_i \sigma_i^2\} & \\ \{_c \mathbf{C}_i^T \sigma_i^2\} & \{_d \mathbf{I}_{c_i} \sigma_i^2\} & \\ \{_c (\boldsymbol{\Sigma}_i \otimes \mathbf{I}_{d_i}) \mathbf{D}_i^T\} & \mathbf{0}^T & \\ & & \{_r \mathbf{D}_i (\boldsymbol{\Sigma}_i \otimes \mathbf{I}_{d_i})\} \\ & & \mathbf{0} \\ & & \{_d \boldsymbol{\Sigma}_i \otimes \mathbf{I}_{d_i}\} \end{bmatrix}.$$

This implies that

$$|\boldsymbol{\Sigma}| = |\{_d \mathbf{I}_{c_i} \sigma_i^2\}| |\{_d \boldsymbol{\Sigma}_i \otimes \mathbf{I}_{d_i}\}| |\mathbf{D}_0 (\boldsymbol{\Sigma}_0 \otimes \mathbf{I}_n) \mathbf{D}_0^T|$$
$$= \left\{ \prod_{i=1}^r \sigma_i^{2c_i} \right\} \left\{ \prod_{i=1}^q |\boldsymbol{\Sigma}_i|^{d_i} \right\} |\boldsymbol{\Sigma}_0|^n,$$

because $\mathbf{D}_0 = \mathbf{I}_{m+1} \otimes \mathbf{I}_n$. The complete data log-likelihood is therefore

$$l = -\frac{1}{2} \sum_{i=1}^r c_i \log \sigma_i^2 - \frac{1}{2} \sum_{i=0}^q d_i \log |\boldsymbol{\Sigma}_i|$$
$$- \frac{1}{2} \sum_{i=1}^r \frac{\mathbf{T}_i^T \mathbf{T}_i}{\sigma_i^2} - \frac{1}{2} \sum_{i=0}^q \mathbf{B}_i^T (\boldsymbol{\Sigma}_i \otimes \mathbf{I}_{d_i})^{-1} \mathbf{B}_i,$$

where

$$\mathbf{B}_0 = \mathbf{Z} - \mathbf{X}\boldsymbol{\beta} - \sum_{i=1}^r \mathbf{C}_i \mathbf{T}_i - \sum_{i=1}^q \mathbf{D}_i \mathbf{B}_i$$

depends on the parameter $\boldsymbol{\beta}$. It follows that the maximum likelihood estimates based on the complete data are

$$\tag{18} \hat{\sigma}_i^2 = \frac{1}{c_i} \mathbf{T}_i^T \mathbf{T}_i, \quad i = 1, \ldots, r,$$

$$\tag{19} \hat{\boldsymbol{\Sigma}}_i = \left[ \frac{1}{d_i} \mathbf{B}_{ij}^T \mathbf{B}_{ik} \right]_{j,k=0}^m, \quad i = 0, \ldots, q,$$

and

$$\tag{20} \mathbf{X}\hat{\boldsymbol{\beta}} = \mathbf{X}[\mathbf{X}^T (\boldsymbol{\Sigma}_0 \otimes \mathbf{I}_n)^{-1} \mathbf{X}]^{-1} \mathbf{X}^T (\boldsymbol{\Sigma}_0 \otimes \mathbf{I}_n)^{-1}$$
$$\cdot \left( \mathbf{Z} - \sum_{i=1}^r \mathbf{C}_i \mathbf{T}_i - \sum_{i=1}^q \mathbf{D}_i \mathbf{B}_i \right).$$

The EM algorithm consists of iteratively replacing $\mathbf{T}_i$ and $\mathbf{B}_i$ in (20), and $\mathbf{T}_i^T \mathbf{T}_i$ and $\mathbf{B}_{ij}^T \mathbf{B}_{ik}$ in (18) and (19), by their conditional expectations given the observed data $\mathbf{Z}$. These expectations are straightforward to calculate because the conditional distributions involved are multivariate normal. Specifically,

$$\mathbf{T}_i | \mathbf{Z} = \mathbf{z} \sim N[\sigma_i^2 \mathbf{C}_i^T \mathbf{V}^{-1} (\mathbf{z} - \mathbf{X}\boldsymbol{\beta}),$$
$$\sigma_i^2 \mathbf{I}_{c_i} - \sigma_i^4 \mathbf{C}_i^T \mathbf{V}^{-1} \mathbf{C}_i],$$

independently, for $i = 1, \ldots, r$, and

$$E(\mathbf{B}_i | \mathbf{Z} = \mathbf{z}) = (\boldsymbol{\Sigma}_i \otimes \mathbf{I}_{d_i}) \mathbf{D}_i^T \mathbf{V}^{-1} (\mathbf{z} - \mathbf{X}\boldsymbol{\beta}),$$
$$\text{var}(\mathbf{B}_i | \mathbf{Z} = \mathbf{z}) = \boldsymbol{\Sigma}_i \otimes \mathbf{I}_{d_i}$$
$$- (\boldsymbol{\Sigma}_i \otimes \mathbf{I}_{d_i}) \mathbf{D}_i^T \mathbf{V}^{-1} \mathbf{D}_i (\boldsymbol{\Sigma}_i \otimes \mathbf{I}_{d_i}),$$

independently, for $i = 1, \ldots, q$.

TABLE 3
*Adjusted means and standard errors for unbalanced apple yield data*

| Treatment | Covariate mean | Response mean | Response Adj.Mean | Std.Err. |
|-----------|----------------|---------------|-------------------|----------|
| A | 8.53 | 283.67 | 269.29 | 13.35 |
| B | 8.40 | 266.67 | 255.69 | 13.35 |
| C | 8.35 | 275.25 | 271.62 | 12.73 |
| D | 7.93 | 270.25 | 277.47 | 12.73 |
| E | 7.48 | 277.25 | 295.96 | 12.73 |
| S | 9.30 | 279.50 | 251.63 | 12.73 |

Adjusted means and their standard errors for the apple yield data from Pearce (1953, 1982), with covariate and response data missing for treatments A and B in block 1. The standard errors were computed using equation (17) and the ML estimates of variance components.

### 5.4 An Unbalanced Example

Consider the apple yield data from Pearce (1953) discussed in Section 2. Suppose that the observations (both covariate and response) were missing for treatments A and B in block number 1. The adjusted means based on this unbalanced data are given in Table 3. The adjusted means are evaluated at the ML estimate of the covariate population mean, $\hat{\mu}_z = 8.2080$, which is not the same as the overall mean covariate value, $\bar{z} = 8.3182$, because of the imbalance with respect to treatments. The standard errors for the adjusted means for treatments A and B are larger than for the other treatments because they are based on observations from three blocks rather than four.

## 6. DISCUSSION

The traditional methods for covariate adjustment of treatment means in designed experiments are inherently conditional. In order to develop a coherent general methodology, we have proposed a multivariate variance components model for the joint distribution of the response and covariates. We have shown that, if the design is orthogonal with respect to blocking factors, then appropriate adjustments to treatment means can be made using the univariate variance components model obtained by conditioning on the observed covariate values. As noted in Section 5, the key to this is the factorization for the joint distribution of $(\mathbf{Y}, \mathbf{Z})$,

$$f(\mathbf{y}, \mathbf{z}; \theta) = f_{Y|Z}(\mathbf{y}|\mathbf{z}; \theta_1) f_Z(\mathbf{z}; \theta_2),$$

where the conditional density $f_{Y|Z}$ defines a univariate linear mixed model for the response variable $Y$,

and where $\theta = (\theta_1, \theta_2)$ and $\theta_1$ and $\theta_2$ are variation independent.

Our approach reveals the fact that some widely used models generate biased adjusted means and incorrect standard errors because the assumed conditional model imposes unrealistic constraints on the joint distribution. Our multivariate model also clarifies some issues that have been the source of long-standing confusion in the statistics literature. One such example is in the analysis of balanced incomplete block designs. As noted by Zelen (1957), "With respect to the non-covariance situation, most statisticians agree that the inter-block analysis may be important if the number of blocks is 'large' or if the variability between blocks is 'small'." However, what is less understood is that the same statement is true in the analysis of covariance. The multivariate analysis makes this clear because it reveals that between block variation in the covariate implies that the slope of the inter-block regression is different from that in the intra-block regression.

In the multivariate model discussed in this paper, we assume that the effect of the covariates is the same for all treatments. It is common in the literature for authors to consider models in which this is not the case. For example, one can easily modify the univariate analysis of covariance model (5), for a randomized blocks experiment, to allow the slope of the covariate regression to depend on the treatment (see, for example, Milliken and Johnson (2002), Chapter 9). However, as we have shown, this univariate analysis is incorrect because it fails to account for the block regression with respect to the covariate. If the block regression components are included in the model, should these also depend on the treatments? It is the opinion of these authors that the correct univariate model for covariate adjustment, if one exists, must be motivated by a multivariate model for the joint distribution of the response and the covariates. For example, a conditional model for the response in which the regression slopes depend on the treatments is implied by a multivariate model in which the error covariance structure is heterogeneous across treatments, but this model also implies that the conditional error variances are heterogeneous across treatments, an assumption that is not typically made. In addition, it seems unnatural to assume heterogeneity in the error covariance structure unless there is also heterogeneity in the block variance–covariance matrices. Thus, it is unclear to these authors if there is a coherent univariate analysis that allows covariate effects to depend on the treatments.

The ideas presented in this paper underscore the importance of proper model specification and careful parameter interpretation in regression analysis of blocked and clustered data. The formulation of the multivariate model guards against ad hoc formulation and misspecification of the regression model by omitting the block-level mean effects that may seriously bias the estimate of the individual-level effects.

A number of articles have explored particular types of adjusting and centering for block and cluster means. There are several reasons for adjusting for the block and cluster means. As noted by Berlin et al. (1999), variability in block and cluster means is common, and can confound the estimated association between the individual-level exposure measurement and outcome; adjusting for the cluster mean may remove confounding bias. Similarly, Neuhaus and Kalbfleisch (1998) argue that inference on the individual-level effects can be misleading without adjustment. Both Kreft, de Leeuw and Aiken (1995) and Raudenbush and Bryk (2002) articulate the need for evaluating block and cluster-level effects as predictor variables in their own right. The paper by Begg and Parides (2003) reviews different heuristic adjustment and centering approaches for the separation of individual-level and block/cluster-level effects on response and their appropriate interpretation. In this paper we suggest a multivariate model that automatically yields the best adjustment and centering suggested by Begg and Parides (2003).

Throughout this article we assumed a joint normal multivariate model. It is well known (see Cambanis, Huang and Simons (1981)) that conditional moment calculations are robust with respect to the family of elliptically contoured distributions. That is, if two random vectors have a joint elliptically contoured distribution, then the conditional distribution of one given the other is also elliptically contoured. The location and scale parameters of the conditional distribution do not depend upon auxiliary parameters of the joint distribution, and consequently, the conditional mean and covariance calculations which apply in the normal case are valid in this more general elliptically contoured setting as well.

In the Bayesian context Gelman (2005) presents a general hierarchical regression approach for ANOVA problems in which effects are structured into exchangeable batches. In this sense, ANOVA is a special case of linear regression, but only if hierarchi-cal models are used. In fact, the batching of effects in a hierarchical model has an exact counterpart in the rows of the analysis of the variance table. In the case where the batches are nonexchangeable Gelman (2005) recommends subtracting batch-level regression predictors, then additive effects for the factor levels in each batch could be modeled as exchangeable. The proposed multivariate variance components model for the joint distribution of the response and covariates would be a better starting point for the hierarchical modeling in the case of nonexchangeable batches. Assigning probability distributions for the treatment effects and variance components automatically leads to coherent Bayesian inferences for the analysis of the covariance model.

Our modeling strategy has assumed, as is traditional in designed experiments, that the covariate values are not affected by the treatments, for example, because they were measured prior to application of the treatments. From a graphical models viewpoint, our model is $B \rightarrow Y \leftarrow Z \leftarrow B$, where $B = (B_y, B_z)$. It is a diversion to try to frame the model in this article on a causal inference scaffold since the inferential goals are quite different. In this article we have outlined a coherent framework for the adjustment of treatment means in designed experiments that account for one or more covariates, whereas in causality one is trying to assess an intervention effect (of $Z$ on $Y$) in the presence of a background variable $(B)$ (Cox and Wermuth (2004)). The commonality of the two issues lies in the fact that in both one is trying to sort out a set of consistent conditional relations within a system of random variables. A goal in casual modeling is to address the overall regression coefficient of $Y$ on $Z$ where $B$ has been decoupled from $Z$, that is, $B$ and $Z$ are nonadjacent in the graph. A consequence of this decoupling is that $\gamma_b$ in (11) equals zero so that the partial and overall effect $Z$ on $Y$ coincide, in which case the conditional model implied by (8) reduces to the univariate mixed model (5). Separating the block effect from the covariate massively restricts the scope of possible models. By starting with a bona fide multivariate model for the joint distribution of response, covariates and blocks, one is led to a sensible univariate conditional model, which properly accounts for the design with respect to the covariate.

Finally, we note that the multivariate variance component model has interesting applications beyond just analysis of covariance. For example, the generalization of a paired $t$-test for a univariate response to multiple observations per subject is a mixed effects model with between and within subject error components. If the response is multivariate, then a multivariate variance components model allows the same generalization to repeated multivariate measurements. The use of multivariate variance components models for repeated measures analysis is considered in Khuri, Mathew and Sinha (1998), Chapter 10.

## APPENDIX: ML ESTIMATION BASED ON THE BIVARIATE MODEL FOR A RCB DESIGN

The representation of the bivariate model given in (9) implies that the joint density of $(\mathbf{y}, \mathbf{z})$ can be factored,

$$f(\mathbf{y}, \mathbf{z}) = \prod_{j=1}^{b} f_{Y|Z}(\mathbf{y}_j|\mathbf{z}_j) f_Z(\mathbf{z}_j).$$

Likelihood-based inference can equivalently be based on the joint density of a 1–1 transformation of the data vector. Specifically, let $\mathbf{H}_t$ denote the Helmet matrix of dimension $t$, and consider the transformation, $(\mathbf{Y}_j, \mathbf{Z}_j) \to (\mathbf{Y}_j^*, \mathbf{Z}_j^*) \equiv (\mathbf{H}^T \mathbf{Y}_j, \mathbf{H}^T \mathbf{Z}_j)$, for $j = 1, \ldots, b$. The $(i, j)$th component of $\mathbf{Y}^*$ is $Y_{ij}^* = \mathbf{h}_i^T \mathbf{Y}_j$, where $\mathbf{h}_i$ is the $i$th column of $\mathbf{H}$. Similarly, $Z_{ij}^* = \mathbf{h}_i^T \mathbf{Z}_j$.

Now, using the facts that $\mathbf{h}_i^T \mathbf{h}_{i'} = 0$ for $i \neq i'$, $\mathbf{h}_i^T \mathbf{h}_i = 1$, and $\mathbf{h}_i^T \mathbf{1} = 0$, for $i = 2, \ldots, t$, it is straightforward to verify that the pairs, $(Y_{ij}^*, Z_{ij}^*)$, $i = 1, \ldots, t$ and $j = 1, \ldots, b$, are mutually independent. Furthermore,

$$\begin{pmatrix} Y_{1j}^* \\ Z_{1j}^* \end{pmatrix} \sim \text{ i.i.d. } N_2 \left[ \begin{pmatrix} \theta_{1,y} \\ \theta_{1,z} \end{pmatrix}, \boldsymbol{\Sigma}_E + t\boldsymbol{\Sigma}_B \right],$$
$$j = 1, \ldots, b,$$

and for each $i = 2, \ldots, t$,

$$\begin{pmatrix} Y_{ij}^* \\ Z_{ij}^* \end{pmatrix} \sim \text{ i.i.d. } N_2 \left[ \begin{pmatrix} \theta_{i,y} \\ 0 \end{pmatrix}, \boldsymbol{\Sigma}_E \right], \quad j = 1, \ldots, b,$$

where $\theta_{i,y} = \mathbf{h}_i^T \boldsymbol{\mu}_y$ and $\theta_{1,z} = \mathbf{h}_1^T \mathbf{1} \mu_z$. It now follows that

$$Z_{1j}^* \sim \text{ i.i.d. } N(\theta_{1,z}, \sigma_{e,z}^2 + t\sigma_{b,z}^2),$$
$$j = 1, \ldots, b,$$
$$Y_{1j}^*|Z_{1j}^* = z_{1j}^* \sim \text{ i.i.d. } N(\theta_{1,yz} + \gamma_{be} z_{1j}^*, \sigma_{be}^2),$$
$$j = 1, \ldots, b,$$
$$Z_{ij}^* \sim \text{ i.i.d. } N(0, \sigma_{e,z}^2),$$
$$i = 2, \ldots, t, j = 1, \ldots, b,$$

and for $i = 2, \ldots, t$,

$$Y_{ij}^*|Z_{ij}^* = z_{ij}^* \sim \text{ i.i.d. } N(\theta_{i,y} + \gamma_e z_{ij}^*, \sigma_e^2),$$
$$j = 1, \ldots, b,$$

where $\gamma_{be} = (\sigma_{e,yz} + t\sigma_{b,yz})/(\sigma_{e,z}^2 + t\sigma_{b,z}^2) = \gamma_b + \gamma_e$, $\theta_{1,yz} = \theta_{1,y} - \gamma_{be}\theta_{1,z}$, and $\sigma_{be}^2 = \sigma_{e,y}^2 + t\sigma_{b,y}^2 - \gamma_{be}(\sigma_{e,yz} + t\sigma_{b,yz})$.

From these distributional results we can easily deduce the ML estimates. In particular, $\hat{\theta}_{1,z} = \bar{z}_1^*$. which implies $\hat{\mu}_z = \bar{z}_{..}$,

$$\hat{\gamma}_{be} = \frac{\sum_{j=1}^{b} (z_{1j}^* - \bar{z}_{1.}^*) y_{1j}^*}{\sum_{j=1}^{b} (z_{1j}^* - \bar{z}_{1.}^*)^2} = \frac{\sum_{j=1}^{b} (\bar{z}_{.j} - \bar{z}_{..}) \bar{y}_{.j}}{\sum_{j=1}^{b} (\bar{z}_{.j} - \bar{z}_{..})^2}$$
$$= \frac{\mathbf{z}^T (\bar{\mathbf{J}}_t \otimes \mathbf{C}_b) \mathbf{y}}{\mathbf{z}^T (\bar{\mathbf{J}}_t \otimes \mathbf{C}_b) \mathbf{z}},$$

$$\hat{\gamma}_e = \frac{\sum_{i=2}^{t} \sum_{j=1}^{b} (z_{ij}^* - \bar{z}_{i.}^*) y_{ij}^*}{\sum_{i=2}^{t} \sum_{j=1}^{b} (z_{ij}^* - \bar{z}_{i.}^*)^2}$$
$$= \frac{\sum_{i=1}^{t} \sum_{j=1}^{b} (z_{ij} - \bar{z}_{i.} - \bar{z}_{.j} + \bar{z}_{..}) y_{ij}}{\sum_{i=1}^{t} \sum_{j=1}^{b} (z_{ij} - \bar{z}_{i.} - \bar{z}_{.j} + \bar{z}_{..})^2}$$
$$= \frac{\mathbf{z}^T (\mathbf{C}_t \otimes \mathbf{C}_b) \mathbf{y}}{\mathbf{z}^T (\mathbf{C}_t \otimes \mathbf{C}_b) \mathbf{z}},$$

$\hat{\theta}_{1,yz} = \bar{y}_{1.}^* - \hat{\gamma}_{be} \bar{z}_{1.}^*$ and $\hat{\theta}_{i,y} = \bar{y}_{i.}^* - \hat{\gamma}_e \bar{z}_{i.}^*$, $i = 2, \ldots, t$. Note that $\hat{\theta}_{i,y} \neq \bar{y}_{i.}^*$. Also, the ML estimate of $\gamma_e$ is identical to the OLS estimate of $\gamma$ based on the standard fixed effects model (1). Also, $\hat{\theta}_{1,y} = \bar{y}_{1.}^* = \mathbf{h}_1^T \bar{\mathbf{y}}_.$, but $\hat{\theta}_{i,y} = \bar{y}_{i.}^* - \hat{\gamma}_e \bar{z}_{i.}^* = \mathbf{h}_i^T (\bar{\mathbf{y}}_. - \hat{\gamma}_e \bar{\mathbf{z}}_.)$, for $i = 2, \ldots, t$. Hence,

$$\hat{\boldsymbol{\mu}}_y = \mathbf{H} \hat{\boldsymbol{\theta}}_y = \mathbf{H} \mathbf{H}^T \bar{\mathbf{y}}_. - \hat{\gamma}_e \mathbf{H} \mathbf{H}^T \bar{\mathbf{z}}_. + \hat{\gamma}_e \mathbf{1} \bar{z}_{..}$$
$$= \bar{\mathbf{y}}_. - \hat{\gamma}_e (\bar{\mathbf{z}}_. - \mathbf{1} \bar{z}_{..}),$$

which agrees exactly with the adjusted mean formula (2) based on the fixed effects model (1).

Finally, if there is no between block variation in the covariate (i.e. $\sigma_{b,z}^2 = 0$), then $\gamma_b = 0$. In this case, $\hat{\gamma}_{be}$ and $\hat{\gamma}_e$ are independent estimates of $\gamma_e$. The ML estimate of $\gamma_e$ in this case is a weighted average of the two independent estimates, with weights inversely proportional to their conditional variances. Since

$$\text{var}(\mathbf{Y}|\mathbf{z}) = (\sigma_e^2 \mathbf{I}_t + \sigma_b^2 \mathbf{J}) \otimes \mathbf{I}_b$$
$$= (\sigma_e^2 + t\sigma_b^2)[(1 - \rho)\mathbf{I}_t + \rho\bar{\mathbf{J}}_t],$$

it follows that

$$\text{var}(\hat{\gamma}_{be}|\mathbf{z}) = \frac{\sigma_e^2 + t\sigma_b^2}{\mathbf{z}^T (\bar{\mathbf{J}}_t \otimes \mathbf{C}_b)\mathbf{z}},$$

and

$$\text{var}(\hat{\gamma}_e | \mathbf{z}) = (1 - \rho) \frac{\sigma_e^2 + t\sigma_b^2}{\mathbf{z}^T (\mathbf{C}_t \otimes \mathbf{C}_b) \mathbf{z}}.$$

This implies that

$$\hat{\gamma}_{e,ML} = \frac{\mathbf{z}^T [(\bar{\mathbf{J}}_t + 1/(1-\rho)\mathbf{C}_t) \otimes \mathbf{C}_b] \mathbf{y}}{\mathbf{z}^T [(\bar{\mathbf{J}}_t + 1/(1-\rho)\mathbf{C}_t) \otimes \mathbf{C}_b] \mathbf{z}}$$

$$= \frac{\mathbf{z}^T [(\mathbf{I}_t - \rho \bar{\mathbf{J}}_t) \otimes \mathbf{C}_b] \mathbf{y}}{\mathbf{z}^T [(\mathbf{I}_t - \rho \bar{\mathbf{J}}_t) \otimes \mathbf{C}_b] \mathbf{z}},$$

which agrees with (7).

## ACKNOWLEDGMENTS

This paper honors the memory of our coauthor Walter T. Federer who died April 14, 2008. Booth supported in part by NSF Grant DMS-08-05865. Wells supported in part by NSF Grant 06-12031 and NIH Grant R01-GM083606-01.